\documentclass{vgtc}                          % final (conference style)
\graphicspath{{figures/}{pictures/}{images/}{./}} % where to search for the images

\usepackage{times}                     % we use Times as the main font
\usepackage{mathptmx}                  % use matching math font

\usepackage{graphicx}  % For \includegraphics and related commands
\usepackage[export]{adjustbox}

\onlineid{0}

\vgtccategory{Research}

\vgtcinsertpkg

\title{Virtual Reality for Immersive Education in Orthopedic Surgery Digital Twins}

\author{Jonas Hein$^{1,2,\circ,}\thanks{email: jonas.hein@inf.ethz.ch}$
\and Jan Grunder$^{1,\circ}$
\and Lilian Calvet$^1$
\and Frédéric Giraud$^1$
\and Nicola Alessandro Cavalcanti$^1$
\and Fabio Carrillo$^{1,\dag}$
\and Philipp Fürnstahl$^{1,\dag}$
}
\affiliation{\scriptsize 
$^1$Balgrist University Hospital, University of Zurich, Zurich, Switzerland\\
$^2$ETH Zurich, Zurich, Switzerland\\
$^\circ$ co-first authors, $^\dag$ co-last authors
}

\teaser{
\centering
\includegraphics[width=0.33\linewidth]{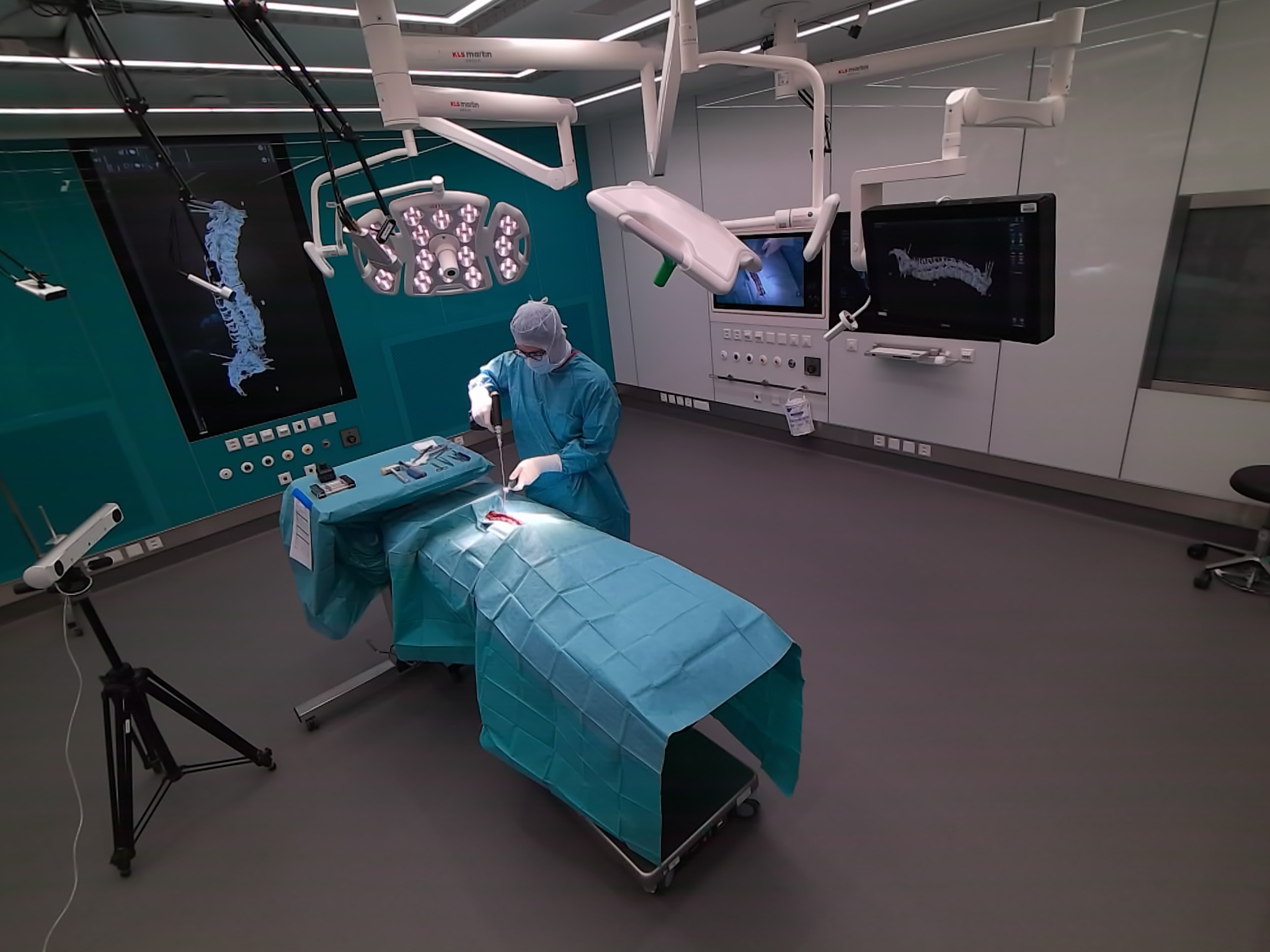}\hfill %
\adjincludegraphics[width=0.33\linewidth, trim={{0.1\width} {0.15\height} {0.0\width} {0.202\height}}, clip, keepaspectratio]{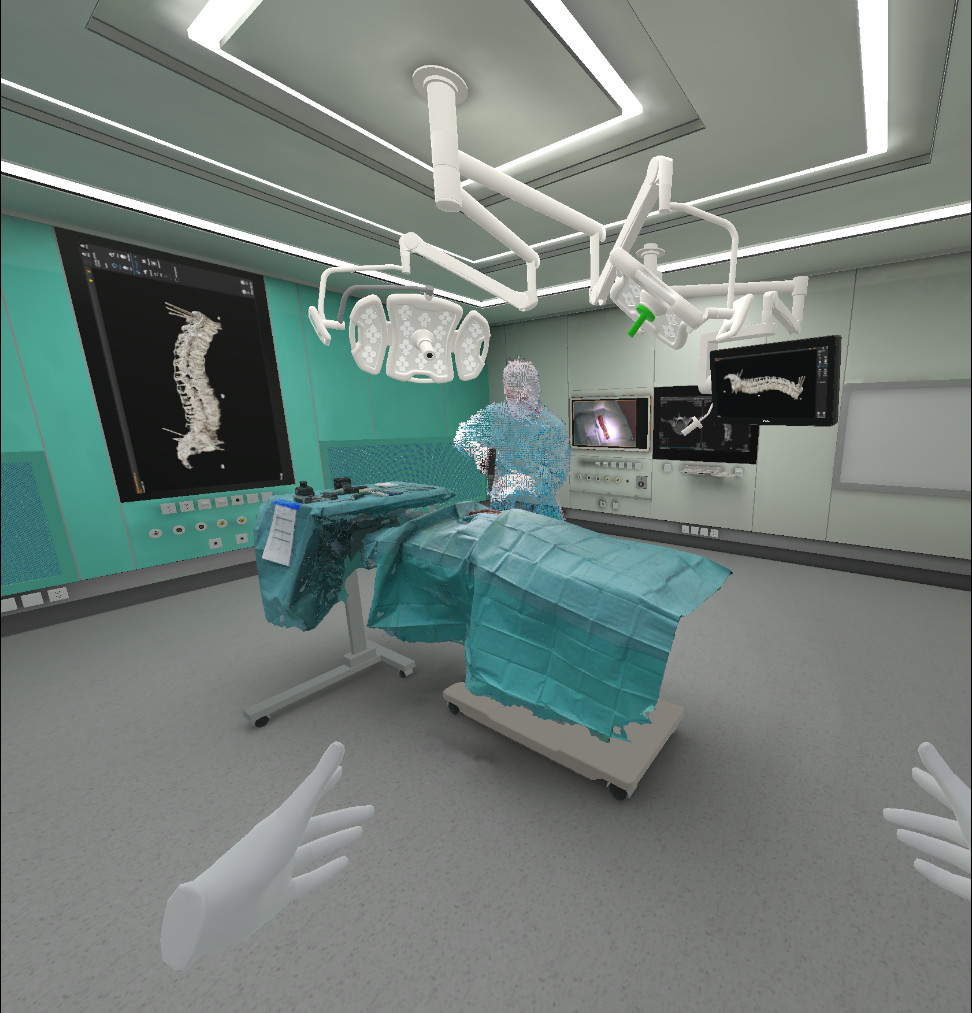}\hfill %
\adjincludegraphics[width=0.33\linewidth, trim={{0.2\width} {0.1\height} {0.0\width} {0.325\height}}, clip, keepaspectratio]{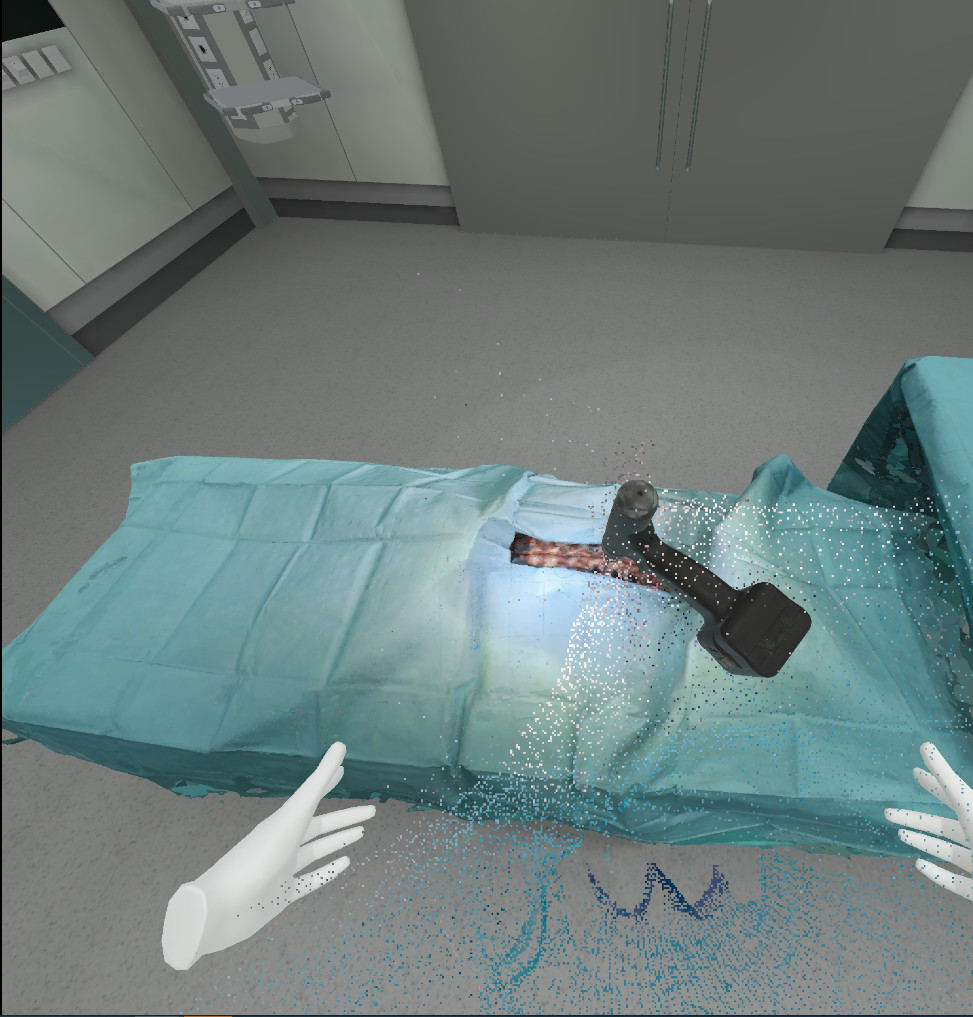}
\caption{
Left: Digital photograph of a spinal surgery captured from a ceiling-mounted camera.
Middle: A user's egocentric view from the proposed VR application, showing the digital replica of the physical surgical scene (digital twin). This includes the operating room, surgical table, anatomy, the surgeon displayed as a dynamic 3D point cloud, and a moving surgical instrument.
Right: Another user's egocentric view from the surgeon's perspective, with a close-up view of the surgical field.
} 
\label{fig:surgery_example}
}

\abstract{
Virtual Reality (VR) technology, when integrated with Surgical Digital Twins (SDTs), offers significant potential in medical training and surgical planning. 
We present SurgTwinVR, a VR application that immerses users within an SDT and enables them to navigate a high-fidelity virtual replica of the surgical environment. 
While several VR and AR systems exist for medical education, ours is the first to propose a virtual dynamic 3D environment that is a clone of a real surgery, encompassing the entire surgical scene, including the surgeon, anatomical structures, surgical instruments, and the operating room. 
Our system is based on the SDT \cite{Hein_2024_CVPR} with some important improvements to make it suitable for real-time rendering, which is essential for an immersive experience. 
It has been enhanced to reduce the time required for displaying the surgeon in action, and additional features have been added to showcase the potential benefits of such an application in surgical education.
} % end of abstract


\begin{document}


\firstsection{Introduction}

\maketitle

% Introduction

The large potential of AR and VR technology for medical education has been explored in numerous studies \cite{curran2023use}.
AR and VR applications have been applied to a wide range of potential use cases such as anatomical education and surgical planning, and emerging evidence suggests that these technologies can provide improved learning outcomes in many scenarios. 
In contrast to conventional teaching materials such as schematics, images or videos, these technologies provide an interactive and more immersive learning experience.
In addition, stereoscopic 3D graphics, as used in most head-mounted displays (HMDs), can provide a more comprehensive spatial understanding of the scene. 
Compared to wet lab training with instructors on real specimens, which is the current gold standard, VR simulations are arguably more ethical, flexible, and significantly cheaper.

Many existing AR and VR applications are based on recorded audio and video material, or on fully synthetic simulation environments.
Recorded video material may be captured from monocular, stereo, or 360° cameras, and optionally enriched with supplementary audio commentary or visual annotations.
While these applications can provide a realistic appearance, the predefined viewpoint is susceptible to obstructions and limits interactions of the user with the scene.
In contrast, VR simulators utilize high-quality 3D models of anatomical structures in fully synthetic environments.
They can provide an interactive and engaging learning experience. However the realism and temporal context of the surgical procedure is often limited.
Furthermore, the 3D models are typically generic and often poorly capture the natural variability of the anatomy or rare pathological cases.
%\textcolor{red}{to be completed with motivations and field of applications.}

In contrast to these existing approaches, we propose to capture and replay real surgeries in 3D.
Our VR application places the user as a virtual participant in a digital replica of the surgical scene, referred to as Surgical Digital Twin (SDT). It enables the user to observe an experienced surgeon performing a surgical procedure and its individual steps. 
The user can freely change their viewing angle, e.g. to avoid occlusions or to take a closer look at instruments or anatomical structures.
Additionally, the user can interact with virtual copies of the surgical instruments used, for example to imitate and internalize the actions of the senior surgeon.
%Similar to normal video players, our application offers the user the most important options for controlling the playback speed, skipping or pausing.
%This allows the user to replay key moments multiple times to focus on different aspects.

The proposed VR application is the first to propose a virtual dynamic 3D environment which is an SDT of an entire surgical scene. 
We propose to demonstrate SurgTwinVR live. 
We believe this demonstration will showcase a very important VR application, and will open discussions for other potential medical applications.

\section{System Overview}
The system involves the creation of the SDT, which includes data collection, fusion, and modeling, as well as the adaptation of the SDT to the requirements of a VR application in terms of immersion and interaction.

\subsection{Creation of the SDT} 
%(approx two third of the column including \cref{fig:camera_positions})
We utilized the digitization approach from \cite{Hein_2024_CVPR} to obtain an SDT of the pedicle drilling step performed during spinal surgery.
The approach focuses on capturing the geometry and appearance of the most important entities in the surgery, namely the anatomy, surgeon, surgical instruments and operating room.
%Since the approach is modular, the SDT comprises a set of independently reconstructed 3D meshes, one for each entity.

\begin{figure}[t]
  \centering
  \adjincludegraphics[width=0.75\linewidth, height=7cm, trim={{0.0\width} {.1\height} {0.0\width} {0.0\height}}, clip, keepaspectratio]{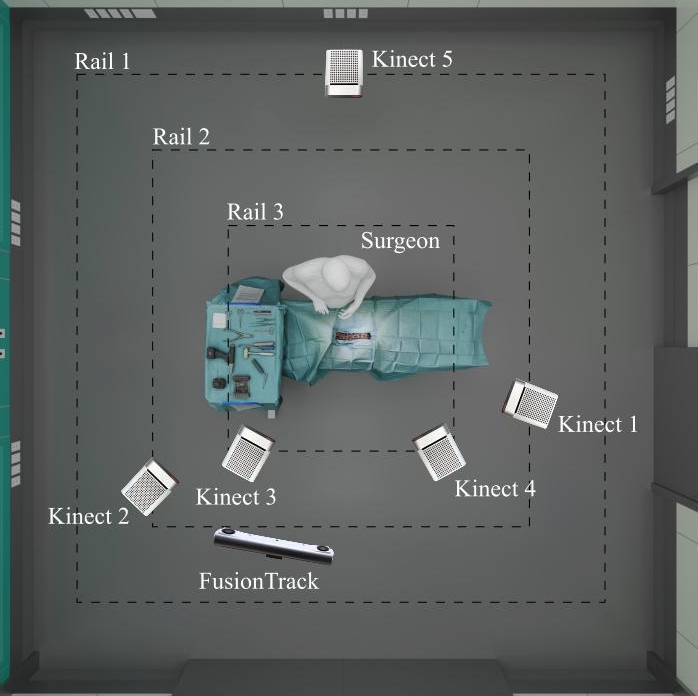}
  \caption{Schematic overview of the SDT's acquisition setup. Five ceiling-mounted \textit{Azure Kinect} RGB-D cameras capture the motion of the surgeon. A \textit{FusionTrack 500} marker-based tracking system captures the trajectory of the surgical instrument.}
  \label{fig:camera_positions}
\end{figure}

The virtual scene is based on a detailed 3D model of the operating room, which was manually designed in an one-time effort, based on a laser scan and detailed photographs.
The model includes permanently installed surgical equipment such as ceiling lamps and displays, and their articulated joints.
In addition, the laser scan served as a reference frame to spatially register all sensors.
The anatomy and operating table were reconstructed using photogrammetry from high-resolution, focus-stacked photographs to ensure an accurate reconstruction of the texture and geometry of the scene.
To dynamically reconstruct the surgeon, RGB-D frames captured with ceiling-mounted cameras (see \textit{Azure Kinect} in \cref{fig:camera_positions}) were fused and converted into a dynamic 3D point cloud representation.
The point cloud was downsampled using voxelization and truncated by a manually defined bounding box of the surgeon's working volume in order to obtain a segmented point cloud representation with uniform density.
The surgical instrument, in our case a surgical drill, was tracked by an optical tracking system (see \textit{FusionTrack} in \cref{fig:camera_positions}) and small infrared-reflective markers attached to the drill.
Both surgeon and instrument were captured at 30 frames per second (FPS).
%An overview of the acquisition setup for surgeon dynamic reconstruction and instrument tracking is shown in Figure \cref{fig:camera_positions}.
%A detailed description of the approach is provided in \cite{Hein_2024_CVPR}.

\subsection{Real-time rendering}
%\todo{@Jan I added a few sentences with some info we should include, but feel free to change/remove anything in here :)}

% The 3D meshes of the operating room and -table and the instruments were down-sampled for a more efficient rendering process while maintaining a realistic appearance.
% We adopt the optimizations from \cite{schütz2021renderingpointcloudscompute} to efficiently render the dynamic point cloud that represents the surgeon.

% Our VR application is implemented based on Unity and deployed on a Meta Quest Pro device.
%While all 3D models and point clouds are transferred to the HMD in advance for this prototype, the dynamic elements could alternatively be streamed to the device in real-time. % this last sentence is wrong. The meta quest link only streams video and not any meshes etc to the HMD

The proposed VR application is implemented in Unity using the Meta XR SDK and the Unity's Universal Render Pipeline.
It runs on a Meta Quest Pro connected to the PC via Link Cable. 
The SDT described in the previous section consists of a scene representation with rigid elements,
namely the operating room, drill, surgical table and anatomy, provided as a set of meshes with PBR materials.
Modifications to the SDT and implementation efforts performed to make it suitable for real-time rendering are concisely described below.

The surgeon is represented as a point cloud rather than a mesh due to the higher geometric accuracy and the holographic effect it offers.
Since Unity does not provide a ready-made solution for dynamic point cloud rendering, a custom compute shader solution was implemented based on the method proposed in \cite{schutz2021rendering}. 
We use two buffers of resolution 1024x1024 pixels which are resolved into the stereo render target. 
This is done as a post processing step after the room geometry has been rasterized. 
In contrast to \cite{schutz2021rendering}, we store color and depth values using only 32 bits, as 64 bit atomic instructions are not supported when running via the Meta Quest Link.
When using 1 byte per color channel and 1 byte for depth, no compression has to occur with the point cloud only having 1 byte per channel as well. 
We mitigate Z-fighting rendering artifacts by storing depth linearly and setting the far clip depth to a low value, which is possible due to the operating room being limited in size.
%To animate the dynamic point cloud at the correct speed, the loading system tracks the time to read the correct animation frame for the current game frame. % important to differentiate between game and animation frames as 1 animation frame will be used for multiple game frames

To minimize rendering times, we downsample the meshes of the reconstructed operating table and the instrument.
The downsampling factor was manually determined for each mesh to find an optimal trade-off between realistic appearance and low rendering time.
%The scanned table and the point cloud of the surgeon from the SDT were downsampled while maintaining a realistic appearance. 
Additionally, we implemented an asynchronous loading mechanism that buffers the current and next point cloud files to reduce loading times for the animation frames.
% TODO talk about performance of the synchronous loading vs asynchronous on average over multiple frames
Compared to simple synchronous loading of animation frames, our implementation reduces loading times by a factor of 10.
%For each game frame the system asynchronously loads the frame immediately after the current animation frame.
%If our animation frame remains as in the last game frame or if it increases by one, the asynchronously loaded data can be used. 
%However if the system lags out it might be that some animation frames need to be skipped. This gets detected and the system loads the correct file in the main thread.
With these performance optimizations, our application runs at an average of 90 FPS on an RTX 3070, which is sufficient to provide a smooth user experience. % will check that again on Monday

\begin{figure}[h]
\centering
% left bottom right top
\adjincludegraphics[width=0.64\linewidth, trim={{0.2\width} {0.15\height} {0.0\width} {0.15\height}}, clip, keepaspectratio]{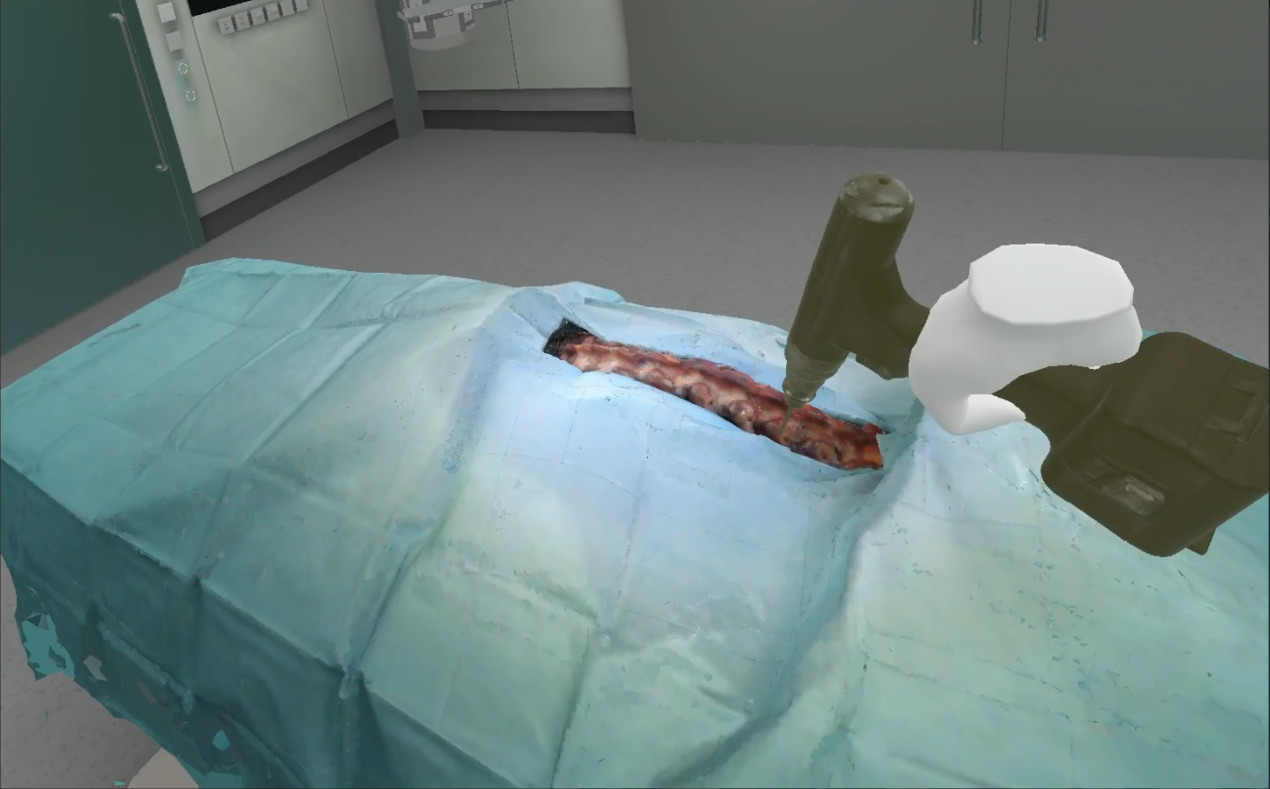}
\caption{
The VR application allows the user to interact with a surgical drill, for example to learn the entry points and angles for pedicle drilling. The playback can be paused and the surgeon and instrument can be hidden temporarily. 
After placing the drill, the user can visually compare their drill pose to the senior surgeons drill pose.} \label{fig:added_feature}
\end{figure}

\section{Conclusion}

We have outlined SurgTwinVR, a VR application for immersive training in orthopedic surgery digital twins.
The system is sufficiently advanced to be demonstrated live. 
This system will draw interest in this important application of VR and will spur discussions about the future use of VR in medical training and surgical planning. 
A first level of interaction with a surgical drill, allowing the user to mimic the gesture of the expert surgeon is proposed (see \cref{fig:added_feature}).
In future work, the proposed experience could be enriched with additional audio commentary or spatial annotations.

\bibliographystyle{abbrv-doi}

\bibliography{main}
\end{document}